\documentclass[preprint,showpacs,preprintnumbers,amsmath,amssymb]{revtex4}
\usepackage{graphicx}% Include figure files
\usepackage{dcolumn}% Align table columns on decimal point
\usepackage{bm}% bold math
\usepackage{epsfig}
\usepackage{subfigure}

\begin{document}

\title{Dynamic Transitions in a Two Dimensional Associating Lattice Gas Model}
\author{Marcia M. Szortyka\footnote[1]{e-mail - marcia.szortyka@ufrgs.br}}
\affiliation{Instituto de F\'{\i}sica, Universidade Federal do Rio Grande do
Sul, Caixa Postal 15051, 91501-970, Porto Alegre, RS, Brazil}

\author{Vera Henriques\footnote[2]{e-mail - vhenriques@if.usp.br}}
\affiliation{Instituto de F\'{\i}sica, Universidade de S\~ao Paulo,
Caixa Postal 66318, 05315970, S\~ao Paulo, SP, Brazil}

\author{Mauricio Girardi\footnote[3]{e-mail - mauricio.girardi@unipampa.edu.br}}
\affiliation{Universidade Federal do Pampa - CP 07 - Bag\'e -
RS - CEP 96400-970
}

\author{Marcia C. Barbosa\footnote[3]{e-mail - marcia.barbosa@ufrgs.br}}
\affiliation{Instituto de F\'isica, Universidade Federal do Rio Grande do
Sul, Caixa Postal 15051, 91501-970, Porto Alegre, RS, Brazil}

\date{\today}
\begin{abstract}
Using Monte Carlo simulations we investigate some new aspects of the phase 
diagram and the behavior of the  diffusion coefficient 
in an associating lattice gas (ALG) model on different regions 
of the phase 
diagram. The ALG model combines a two dimensional lattice gas 
where particles interact 
 through  a soft core potential and orientational degrees of 
freedom. The competition 
between soft core potential and directional attractive 
forces results in a high density liquid phase, 
a low density liquid phase, and a gas phase. Besides anomalies in 
the behavior of the
 density with the temperature at constant pressure and of
the  diffusion coefficient with density at constant temperature
are also found. 
The two liquid phases are separated by a coexistence
line that ends in a bicritical point. The low density 
liquid phase is separated from the gas phase by a coexistence line
that ends in  tricritical point. The
bicritical and tricritical points are linked by a critical
$\lambda$-line. The high density liquid phase
and the fluid phases are separated by a second $\tau$
critical line. We then investigate how the diffusion coefficient behaves on 
different 
regions of the chemical potential-temperature phase diagram. We find 
that diffusivity undergoes two types of dynamic transitions: a fragile-to-strong  transition when the  critical $\lambda$-line is crossed by decreasing
the temperature at a 
constant chemical potential; and
a  strong-to-strong transition when  the $\tau$-critical line
is crossed
by decreasing the temperature at a constant   chemical potential.

\end{abstract}

\maketitle
%%%%%%%%%%%%%%%%%%%%%%%%%%%%%%%%%%%%%%%%%
%%%%%%%%%%%%%%%%%%%%%%%%%%%%%%%%%%%%
\section{\label{sec1}Introduction}
%%%%%%%%%%%%%%%%%%%%%%%%%%%%%%%%%%%%
%%%%%%%%%%%%%%%%%%%%%%%%%%%%%%%%%%%%%%%%%
The study of the  properties of supercooled water
is motivated by its well known anomalous thermodynamic behavior.
Besides the density anomaly, the response functions for water
appear to diverge at a singular 
temperature $T_c=228\;K$ \cite{Sp76}. 
This  apparent divergence of the response functions led to 
the hypotheses of the existence of liquid polimorphism
and of a second critical point, at $T_c=228\;K$  \cite{Po92}.
In spite of the enormous attention given to 
this possible singularity, as well as to the many other anomalies, no unique explanation
has yet been established.
The hypothetical singular point is hidden below the 
homogeneous nucleation temperature $T_H=235\;K$ \cite{Ra72} in an
experimentally inaccessible temperature range for bulk supercooled water.
This rules out direct experimental investigation of this region in
order to confirm the existence of liquid-liquid coexistence.
In order to circumvent this difficulty, it has been proposed, recently,
that a dynamic crossover  of the transport properties such
as the self-diffusion constant, $D$, and the viscosity, $\eta$, at temperatures
above $T_c$, would indicate the 
 presence of a critical point \cite{Li05}\cite{Ch06}.
The dynamic crossover has also been associated with liquid-liquid
transitions in silicon \cite{Sa03} and 
in non-tetrahedral liquids \cite{Xu06b}.

The basic surmise behind the link between the dynamic crossover and
the presence of a second critical point goes as follows.
The liquid-liquid coexistence line that separates two liquid phases
terminates  at a critical
point. Beyond this point, at which the response functions diverge, one finds
lines of maxima of these functions which assymptotically approach the critical point.
This extension of the first-order phase boundary into
the one-phase region 
is the Widom line at $T_L(P)$.   Even though this line does not 
exhibit any thermodynamic transition, experiments on  water show that
the specific heat, shear viscosity and thermal diffusivity \cite{An04}
exhibit a peak when crossing the Widom line. In particular, Maruyama et. al 
\cite{Ma04} conducted experiments in nanopores (to avoid homogeneous
nucleation) at 
ambient pressure that present a peak at the constant pressure specific heat 
at $T_{Cp}=227\;K$. This temperature coincides (within the 
experimental
error bar)  with that one 
temperature obtained by Xu et. al \cite{Xu05}, $T_{Cp}=225\;K$  for the 
location of
the dynamic crossover  suggesting that 
this crossover occurs at the Widom line, confirming  the
presence of the second critical point.

Unfortunately, the presence of 
a peak in the specific heat in a certain region of the
pressure-temperature phase-diagram is not exclusivity of  Widom lines. 
For instance, in glassformers an abrupt heat capacity drop
is observed  when ergodicity is broken. 
This change can happen very sharply in the case of fragile liquids 
or it may take tens of degrees in the case of strong liquids.
Examples of fragile liquids are toluene and metallic systems, while
covalent and network forming systems are strong liquids \cite{An95}.
In the last case, the increase in the specific heat can be 
simply a smeared peak, located above the melting temperature,$T_m$, like
in the case of $SiO_2$ and of 
$BeF_2$
\cite{Ri82a}\cite{Ri82b}\cite{Ta75a}\cite{Ta75b}\cite{Sc05}
\cite{So27}. In the case of strong liquids, it is also possible 
to observe a weak transition
 at a temperature between the glass transition temperature, $T_g$, and 
the melting temperature, $T_m$. This peak in the
specific heat curve occurs
in the tail of a $\lambda$ thermodynamic transition where
there is a little heat capacity to loose \cite{An08}. Example
of such strong liquids are the tetrahedral bonded liquids such as 
water, Si and Ge. This implies that observing a fragile-to-strong 
crossover in a region where the specific heat grows does not
univocally imply the presence of a critical point. 
An interesting question, however, would be: does the presence
of criticality result in a fragile-to-strong crossover?

In order to address this point, in this paper
we analyze a model that 
exhibits two different critical lines and 
we explore what happens with the
dynamics close to these line, in order to test
if a fragil-to-strong transition would
be a signature for criticality. 
The present model is an Associating Lattice Gas (Henriques and Barbosa)
that corresponds to a lattice gas with  hydrogen 
bonds representede
 through ice variables. A competition between the filling
up of the lattice and the formation of an open four-bonded orientational 
structure is naturally
introduced in terms of the ice bonding variables, and {\it no ad hoc 
} addition of density or
bond strength variations is needed. Besides the gas phase and as a result of 
this competition, the model exhibits two liquid phases that
bare resemblance to the two liquid phases 
predicted for water, corresponding
to  a low
density liquid phase and a high density 
liquid phase. Moreover, it has both the diffusion and the
density anomalies present in water \cite{Sz07}. 
 
Here,  the model phase diagram is reviewed and analyzed  for the presence
of dynamic transitions. 
Two new critical lines were found beyond the liquid-liquid coexsitence line. 
We searched for fragile to strong transitions in the proximity of these
two lines.  
Comparison between the behaviors of the specific heat and of the 
diffusion constant in these regions may help
in understanding  if the type of dynamic transition observed
in confined water necessarily means the presence of criticality.

The remaining of this article goes as follows. In sec. II, the 
lattice model is reviewed, for clarity. 
In sec. III, results for the chemical potential-temperature phase-diagram 
are shown and discussed. Our investigation of
diffusion is presented in sec. IV.  Sec V is a final section of conclusions.

%%%%%%%%%%%%%%%%%%%%%%%%%%%%%%%%%%
%%%%%%%%%%%%%%%%%%%%%%%%%%%%%%%%%%
\section{\label{sec2} The Model}
%%%%%%%%%%%%%%%%%%%%%%%%%%%%%%%%%%
%%%%%%%%%%%%%%%%%%%%%%%%%%%%%%%%%%

\begin{figure}
\begin{centering}
\includegraphics[scale=0.25]{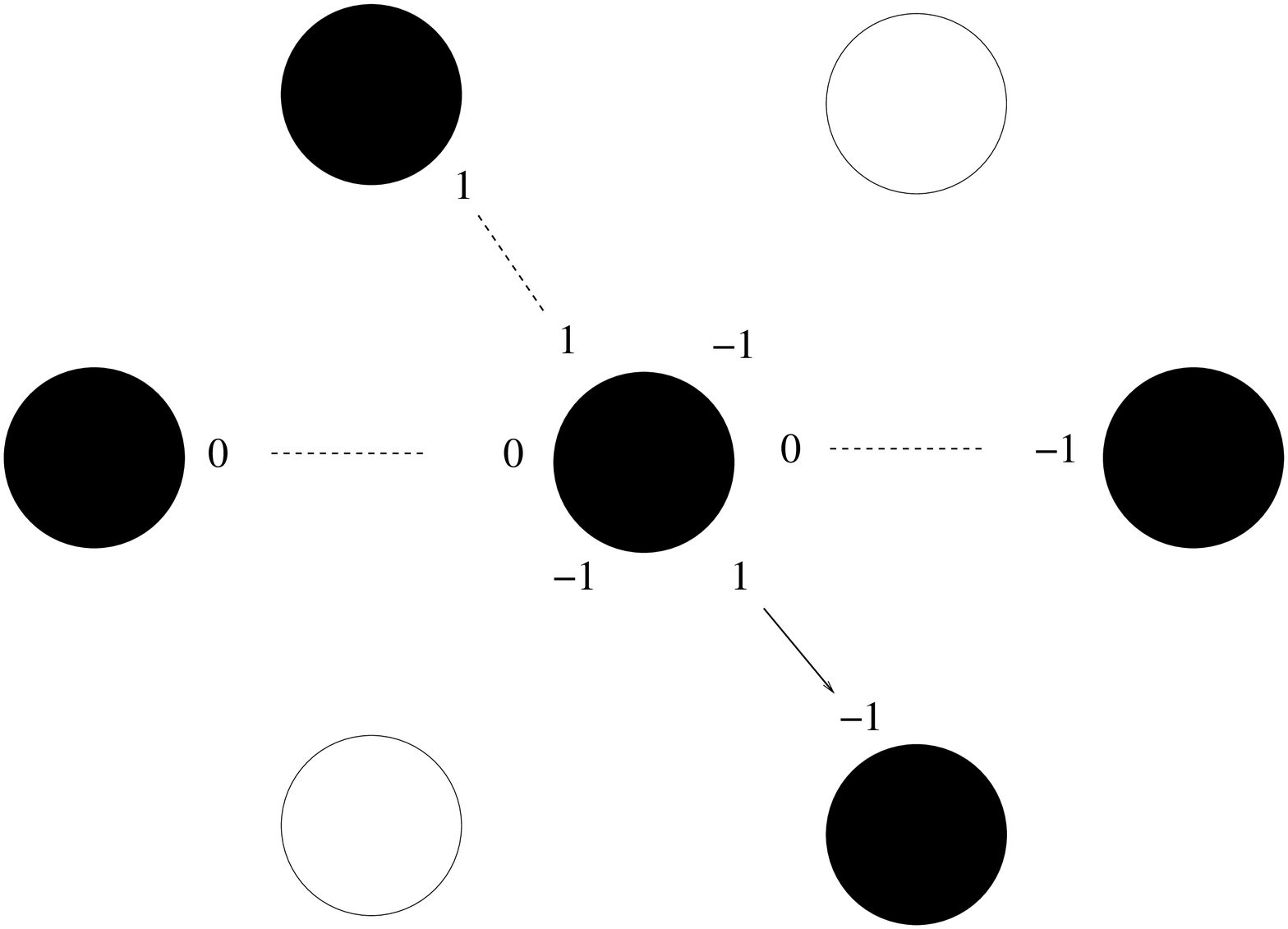} 
\par\end{centering}
\caption{Particles in the model: An occupied central site $i$ and its
 six bond variables, $\tau_{\scriptscriptstyle{i}}^{A}$, with $A=1,..,6$. If 
$\tau_{\scriptscriptstyle{i}}^{A}=0$
no bond is formed in spite of the configuration of the arm of the
neighbor site. If $\tau_{\scriptscriptstyle{i}}^{A}=\pm1$ and the neighbor's arm 
is $\tau_{\scriptscriptstyle{k}}^{B}=\mp1$, a bond is formed. Dashed lines 
represent a non bonding 
configuration, while the solid line represents a bonding configuration.}
\label{part} 
\end{figure}

We consider a two-dimensional lattice gas model of size $L^2$ on a 
triangular lattice
as introduced by \emph{Henriques} and \emph{Barbosa} \cite{He05a}. 
In this model, particles are represented by an occupational variable,
$\sigma_{\scriptscriptstyle{i}}$, which assumes the value 
$\sigma_{\scriptscriptstyle{i}}=0$, if the site
is empty, or $\sigma_{\scriptscriptstyle{i}}=1$, if the site is 
full, and six orientational variables, 
$\tau_{\scriptscriptstyle{i}}^{A}$, 
that represent the different orientations that the particle might exhibit. If
two neighboring sites have complementary orientations, a hydrogen bond is 
formed.
Four bonding variables are the ice bonding arms: two donors, with 
$\tau_{\scriptscriptstyle{i}}^{A}=1$,
and two acceptors, with $\tau_{\scriptscriptstyle{i}}^{A}=-1$. The 
other two arms, 
with $\tau_{\scriptscriptstyle{i}}^{A}=0$, do not form bonds, and are taken always
opposite to
each other, as illustrated in Fig.(\ref{part}). There is no restriction 
for donor/acceptor 
arms positions, thus there are eighteen possible states for each occupied site. 

The Hamiltonian includes two contributions: an isotropic, 
van der Waals like interaction, while the second interaction depends on the
orientational degrees of freedom. Two neighboring 
sites, $i$ and $k$, with pointing arms $A$ and $B$, 
form a hydrogen bond if the product between their
orientational variables is given by
$\tau_{\scriptscriptstyle{i}}^{A}\tau_{\scriptscriptstyle{k}}^{B}=-1$, 
yielding an energy per site $e = E/L^2 = -v$. For a non-bonding pair of 
occupied sites, 
the energy per site is $e = -v+2u$, for $u>0$.
In spite of the fact that each molecule may have six neighbors, only four hydrogen bonds
per particle are allowed. 
The overall energy of the system is given by 

%%%%%%%%%%%%%%%%%%%%%%%%%%%%%%%%%%%%%%%%%%%%%%%%%%%%%%%%%%%%%
\begin{equation}
\mathcal{H}=\left(-v+2u\right)\sum_{\left\langle i,k\right\rangle}
\sigma_{\scriptscriptstyle{i}}\sigma_{\scriptscriptstyle{k}}+
u\sum_{\left\langle i,k\right\rangle}
\sigma_{\scriptscriptstyle{i}}\sigma_{\scriptscriptstyle{k}}
\left[\left(1-\tau_{\scriptscriptstyle{i}}^{A}
\tau_{\scriptscriptstyle{k}}^{B}\right)
\tau_{\scriptscriptstyle{i}}^{A}\tau_{\scriptscriptstyle{k}}^{B}\right]
\;\;,\label{E}
\end{equation}
%%%%%%%%%%%%%%%%%%%%%%%%%%%%%%%%%%%%%%%%%%%%%%%%%%%%%%%%%%%%%
 where $\sigma_{\scriptscriptstyle{i}}=0,1$ are occupation variables,
$\tau_{\scriptscriptstyle{i}}^{A}=0,\pm1$ represent the arm state variables, 
the summation $\langle i,k\rangle$ is over neighboring sites.

%%%%%%%%%%%%%%%%%%%%%%%%%%%%%%%%%%%%%%%%%%%%%%%%%%%%%%%%%%%%%
%%%%%%%%%%%%%%%%%%%%%%%%%%%%%%%%%%%%%%%%%%%%%%%%%%%%%%%%%%%%%

\begin{figure}
\begin{centering}
\includegraphics[scale=0.35]{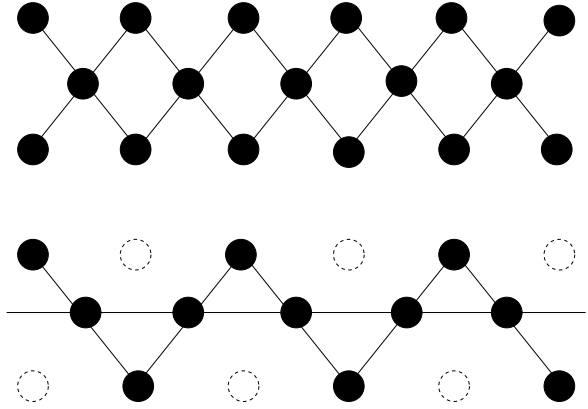} 
\par\end{centering}
\caption{In the high density phase, {\bf HDL}, the lattice is full,
and an energy punishment arises, because 
two inert arms point to filled sites. In the 
low density phase, {\bf LDL}, the lattice is 
3/4 filled and particles are distributed over the lattice
in such a way that the inert arms point only to the 
empty sites. There is no energy punishment, in this case.}
\label{phases} 
\end{figure}

Comparing the energies of the model at zero temperature two liquid phases, a 
low density (LDL) and 
high density  (HDL) liquid phase are found, besides the gas phase. Fig.(\ref{phases}) illustrates 
the HDL and LDL phases. 
For high  values of the chemical potential the lattice is fully occupied (
density $\rho=1$ )
and the energy per site is
$e  = -3v+2u$. At lower values of the chemical potential, $\mu$, the soft core 
repulsion becomes 
dominant, and the lattice 
becomes $3/4$ filled, with density $\rho=0.75$ and 
energy per site 
$e=-\frac{3}{2}v$. Like every other lattice gas, 
the model exhibits a gas phase, at very low chemical potentials.

At zero temperature, the grand potential per site, $\phi = \Phi/L^2$, 
is given by
%%%%%%%%%%%%%%%%%%%%%%%%%%%%%%%%%%%%%%%%%%%%%%%%%%%%%%%%%%%%%
\begin{equation}
\phi(T=0) = \langle \mathcal{H}-\mu\sum_{i}\sigma_{\scriptscriptstyle{i}}
\rangle = E - \mu N \;\;.\label{pot}
\end{equation}
%%%%%%%%%%%%%%%%%%%%%%%%%%%%%%%%%%%%%%%%%%%%%%%%%%%%%%%%%%%%%
By equating the grand potential of different phases, we find that the high density 
phase (HDL) coexists with the low density 
phase (LDL)  at  the reduced chemical potential $\overline{\mu}=\mu/v = -6 + 8u/v$. The 
coexistence between the LDL and the gas phases
occurs at $\mu/v = -2$.

%%%%%%%%%%%%%%%%%%%%%%%%%%%%%%%%%%%%%%%%%%%%%%%%%%%%%%%%%%%%%
%%%%%%%%%%%%%%%%%%%%%%%%%%%%%%%%%%%%%%%%%%%%%%%%%%%%%%%%%%%%%
The properties of the
system at finite temperatures were obtained from
Monte Carlo simulations in the grand canonical ensemble, through the
Metropolis algorithm. 
We present a detailed study of the model system, for $L = 30$. Some finite size scaling analysis
was also undertaken, when necessary. Interaction parameters 
were fixed at $u/v = 1$, which corresponds to "repulsive" van der Waals interaction.
Reduced parameters are defined by
%%%%%%%%%%%%%%%%%%%%%%%%%%%%%%%%%%%%%%%%%%%%%%%
\begin{eqnarray} 
\overline{p} &=& \frac{p}{v} \;\;  \nonumber \\
\overline{T} &=& \frac{k_B\;T}{v}\;\; \\
\overline{\mu} &=& \frac{\mu}{v}. \nonumber \\
\label{par}
\end{eqnarray}
%%%%%%%%%%%%%%%%%%%%%%%%%%%%%%%%%%%%%%%%%%%%
Equilibrium transitions were investigated through
analysis of the system specific heat.
First-order transitions point were located from hysteresis.
The constant volume specific heat was calculated from simulation data
obtained at constant chemical potential through the expression 
%%%%%%%%%%%%%%%%%%%%%%%%%%%%%%%%%%%%%%%%%%%%%%%
\begin{eqnarray} 
c_V=\frac{1}{k_B T^2 V}( \langle \delta \phi^2 \rangle_{\mu V T} - \frac{\langle \delta\phi \delta \rho \rangle^2_{\mu V T}}{\langle \delta \rho^2 \rangle_{\mu V T} })
\label{CV}
\end{eqnarray}
%%%%%%%%%%%%%%%%%%%%%%%%%%%%%%%%%%%%%%%%%%%%
 adapted from \cite{Al87} to the lattice. $\rho$ is  the density, $V$ is the volume
 and $\delta X= X - \langle X \rangle$ with $X=\phi,\rho$.

%%%%%%%%%%%%%%%%%%%%%%%%%%%%%%%%%%%%%%%%%
%%%%%%%%%%%%%%%%%%%%%%%%%%%%%%%%%%%%%%%%%
\section{\label{sec3} The Phase Diagram}
%%%%%%%%%%%%%%%%%%%%%%%%%%%%%%%%%%%%%%%%%
%%%%%%%%%%%%%%%%%%%%%%%%%%%%%%%%%%%%%%%%%

The chemical potential-temperature phase diagram of the model 
was partially analyzed in previous work \cite{He05a}, which focused on
the coexistence lines between the low and high
density liquids. In this paper the $\overline{\mu}-\overline{T}$ phase-diagram is 
complemented by the analysis of the region beyond the coexistence line.

\begin{figure}
\begin{centering}
\includegraphics[clip=true, scale=0.5]{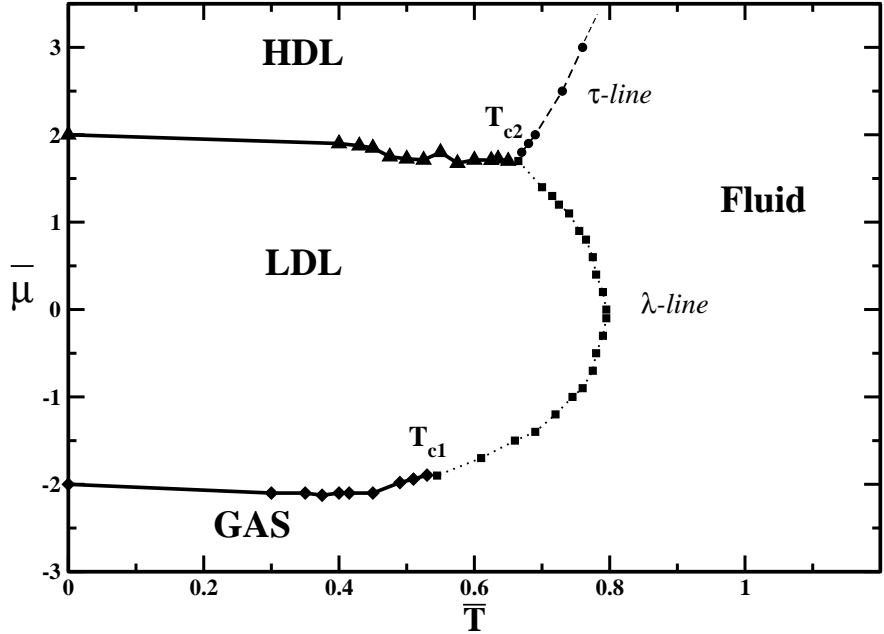}
\par\end{centering}
\caption{Phase diagram showing reduced chemical vs. reduced temperature.
The diamonds represent the Gas-LDL coexistence line. 
The triangles indicate the LDL-HDL coexistence line. $T_{c1}$ is the tricritical  point Gas-LDL and $T_{c2}$ is the tricritical point LDL-HDL.
 The squares and circles are lines, obtained by the maximum 
in specific heat, that separates fluid phase from 
LDL and HDL phases, respectively. The zero temperature 
points, at $\overline{\mu}=-2$ and $\overline{\mu}=2$, are exact.}
\label{mu-t}
\end{figure}

The complete $\overline{\mu}-\overline{T}$ phase-diagram is illustrated in Fig.(\ref{mu-t})
and goes as follows. At low reduced chemical potentials, $\overline{\mu}$, 
for all reduced temperatures, $\overline{T}$
only the gas phase is present. As the reduced chemical potential increases
a low density liquid phase appears. This phase coexists with
the gas phase along a first-order transition line at $\overline{\mu}=\overline{\mu}_{gas-LDL}(\overline{T})$.
For even higher reduced
chemical potentials a high density liquid phase emerges. This
phase coexists with the low density liquid phase at the 
first-order line $\overline{\mu}=\overline{\mu}_{LDL-HDL}(\overline{T})$.
 But what happens at the end of the two first-order lines? 

\begin{figure}
\begin{centering}
\vspace{0.8cm}
\includegraphics[clip=true,scale=0.5]{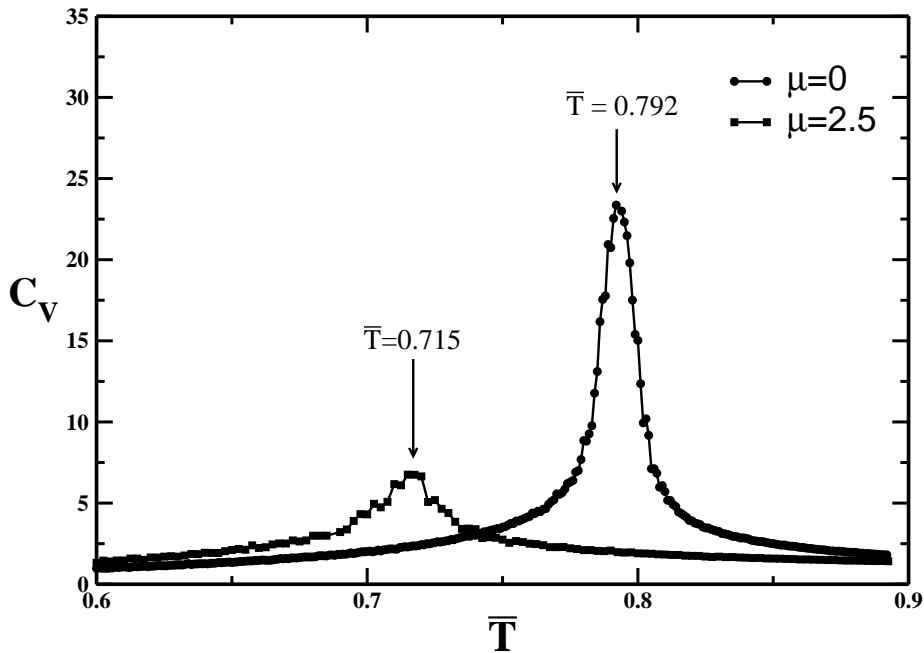}
\par\end{centering}
\caption{ Specific heat at constant volume for 
$\overline{\mu}=0$ 
and $\overline{\mu}=2.5$.}
   \label{cv}
\end{figure}

In order
to answer the question, we have examined the specific heat at
 constant volume, 
$c_{V}$, as a function
of temperature, for fixed values of $\overline{\mu}$, in
two regions of the $\overline{\mu}-\overline{T}$ phase diagram: 
between the two coexistence lines and above the LDL-HDL coexistence line.
Fig.(\ref{cv}) illustrates our results.
 For  $\overline{\mu} =0$,  between the two coexistence lines, $c_V$ has a peak
at a reduced temperature 
$\overline{T}=\overline{T}_{\lambda}\approx 0.79$, suggesting
the presence of criticality.  Similar behavior was observed for every investigated
chemical potential between the two coexistence lines, 
indicating the presence of a critical line. We called this line $\lambda$ and represented 
it in Fig.(\ref{mu-t}) through a dotted line and square symbols.  Above the liquid-liquid
coexistence line, for
 $\overline{\mu} = 2.5$, the specific heat, $c_V$, displays  also a peak
at $\overline{T}=\overline{T}_{\tau}\approx 0.71$. We have examined a range of
chemical potentials above the LDL-HDL coexistence line. 
A line of maxima of these  peaks, named $\tau$ was added to the $\overline{\mu}-\overline{T}$ phase diagram,
as shown in in Fig.(\ref{mu-t}) (dashed line and circles).

\begin{figure}
\begin{centering}
 \includegraphics[clip=true,scale=0.5]{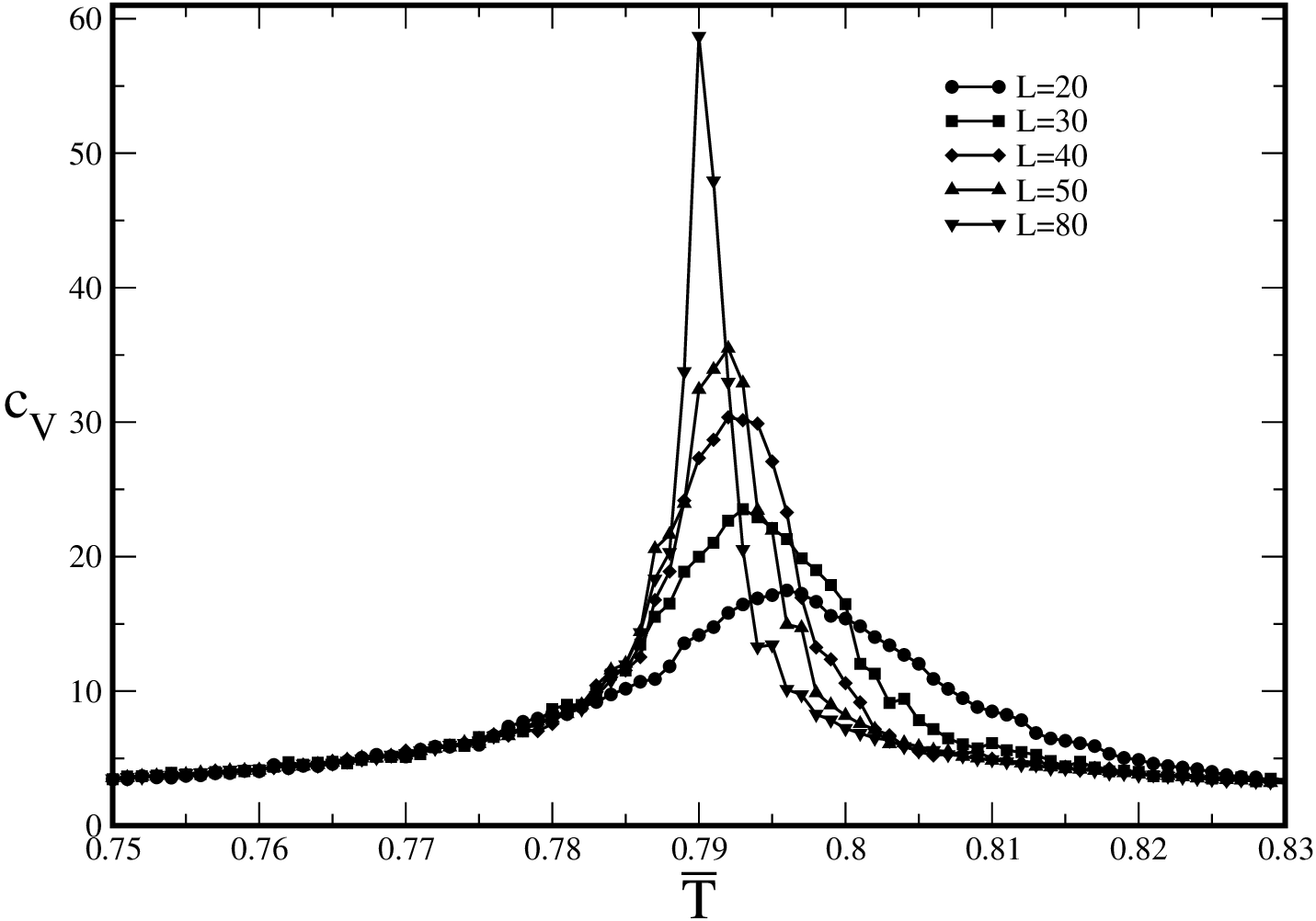}
 \par\end{centering}
\caption{
Specific heat at constant volume for different 
lattice sizes in the 
 fluid/LDL region ($\overline{\mu}=0$).}
\label{cv_L_lambda}
\end{figure}

\begin{figure}
\begin{centering}
 \includegraphics[clip=true,scale=0.5]{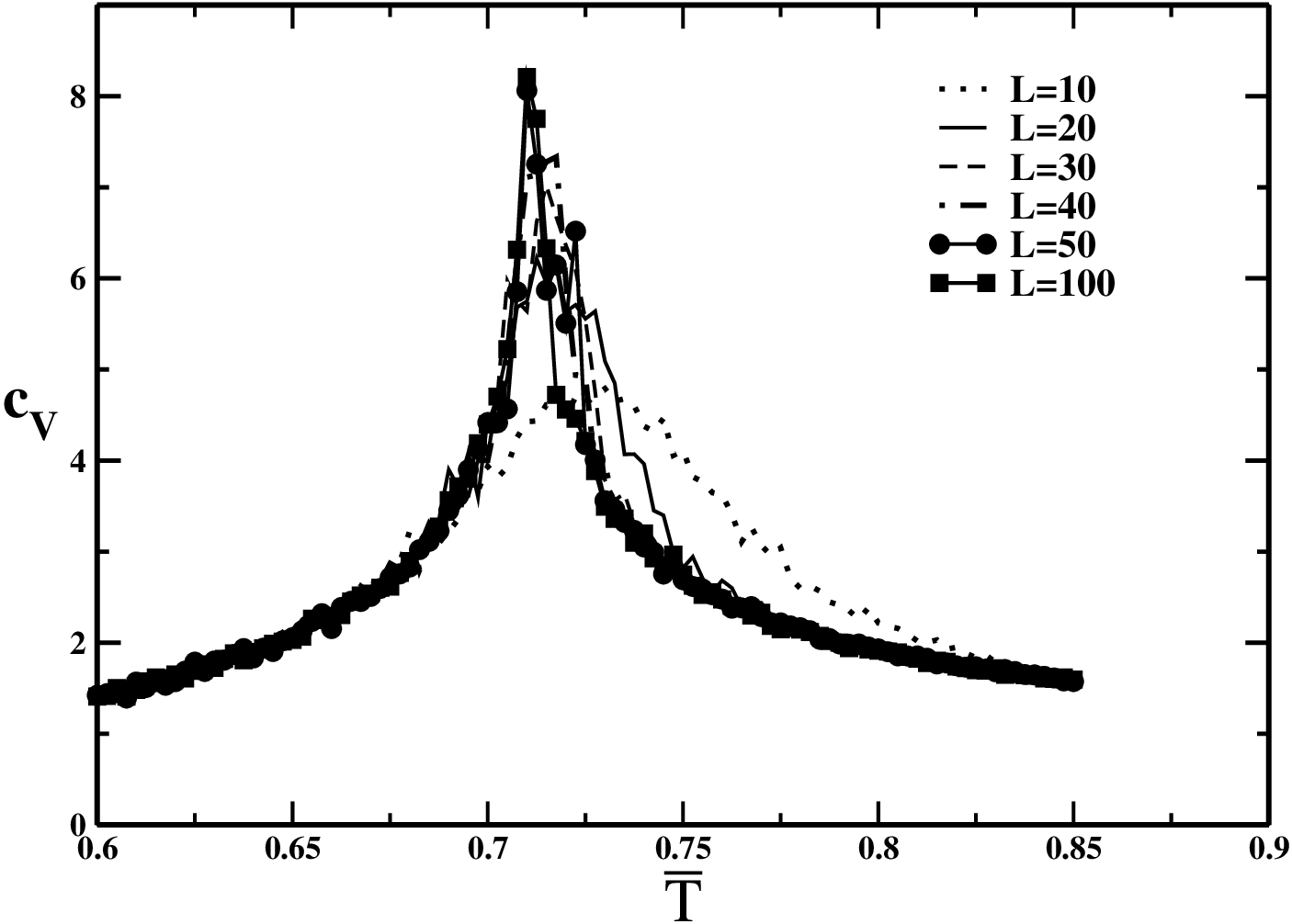}
\par\end{centering}
\caption{
 Specific heat at constant volume for different 
lattice sizes in the 
 fluid/HDL region ($\overline{\mu}=2.5$).}
     \label{cv_L_tau}
\end{figure}

In order to check the nature of the two lines,
$\lambda$ and $\tau$, in
Fig. (\ref{mu-t}),  the specific heat $c_V$  
within these regions was computed for different lattice sizes
(L=10, 20, 30, 40, 50, 80 and 100).
Fig.(\ref{cv_L_lambda})  illustrates the behavior of $c_V$
for $\overline{\mu}=0$ for various lattice sizes, showing
a diverging peak, as $L\rightarrow \infty$.  
 Fig.(\ref{cv_L_tau}) shows $c_V$ for $\overline{\mu}=2.5$, for
 various lattice sizes. In this case, however, the 
peak increases mildly with $L$.

\begin{figure}
\begin{centering}
 \includegraphics[clip=true,scale=0.5]{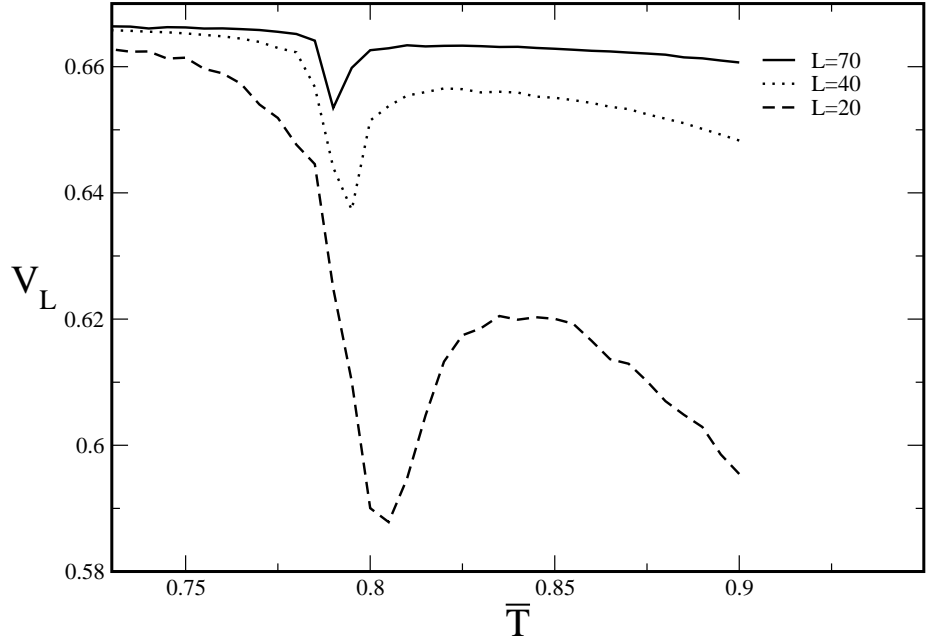}
 \par\end{centering}
\caption{Energy cumulant for the  
 fluid/LDL region ($\overline{\mu}=0$).}
\label{cumulant_lambda}
\end{figure}

\begin{figure}
\begin{centering}
 \includegraphics[clip=true,scale=0.5]{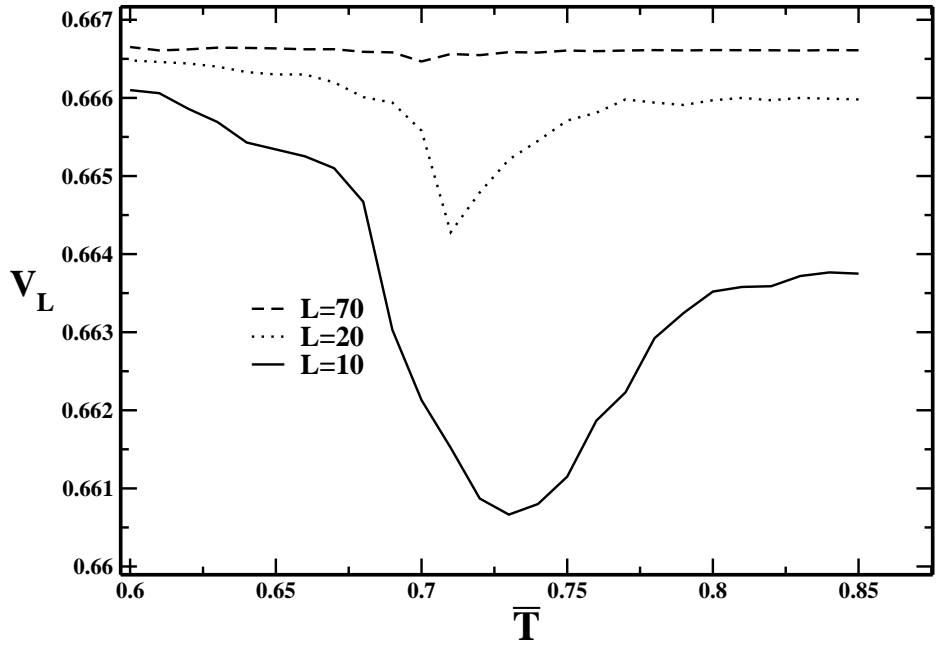}
\par\end{centering}
\caption{
 Energy cumulant for the 
 fluid/HDL region ($\overline{\mu}=2.5$).}
     \label{cumulant_tau}
\end{figure}

The criticality of $\lambda$ and $\tau$ was investigated by calculating 
the energy cumulant given by
%%%%%%%%%%%%%%%%%%%%%%%%%%%%%%%%%%%%%%%%%%%%%%%%%%%%%%%%%%%%%
\begin{equation}
V_L=1-\frac{\langle E^4 \rangle}{3\langle E^2\rangle^2}\;.
\label{cumulant}
\end{equation}
%%%%%%%%%%%%%%%%%%%%%%%%%%%%%%%%%%%%%%%%%%%%%%%%%%%%%%%%%%%%%
Fig.(\ref{cumulant_lambda}) illustrates the 
energy cumulant for $\mu=0$, showing the signature
for criticality. Fig.(\ref{cumulant_tau}) illustrates the
energy cumulant for $\mu=2.5$ that also indicates
the presence of criticality.

\begin{figure}
\begin{centering}
\includegraphics[clip=true,scale=0.5]{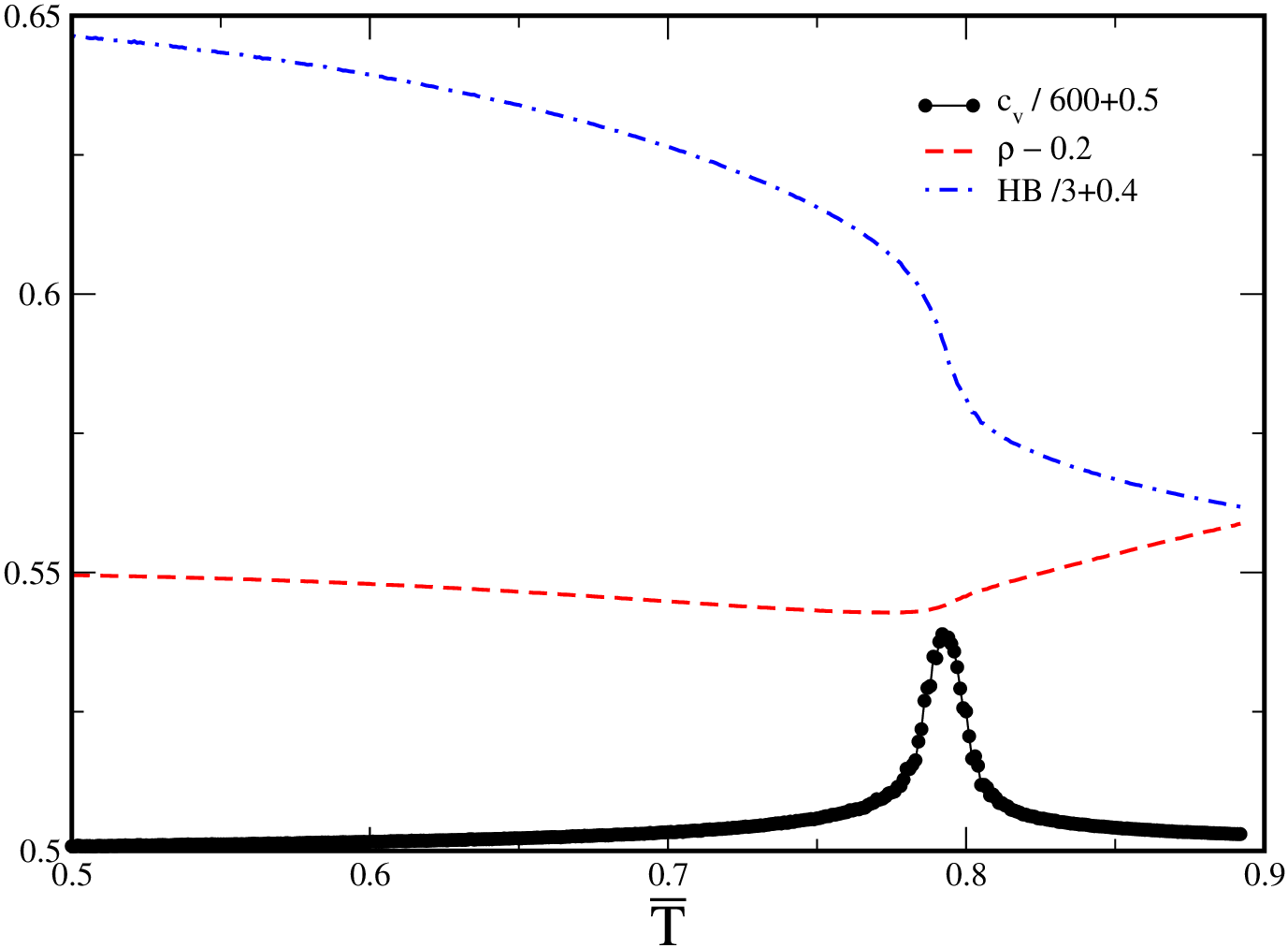}
\par\end{centering}
\caption{(Color on line) 
Density, specific heat  and 
number of bonds per site versus reduced temperature for $\overline{\mu}=0$} 
\label{dw_lambda}
\end{figure}

\begin{figure}
\begin{centering}
\includegraphics[clip=true,scale=0.5]{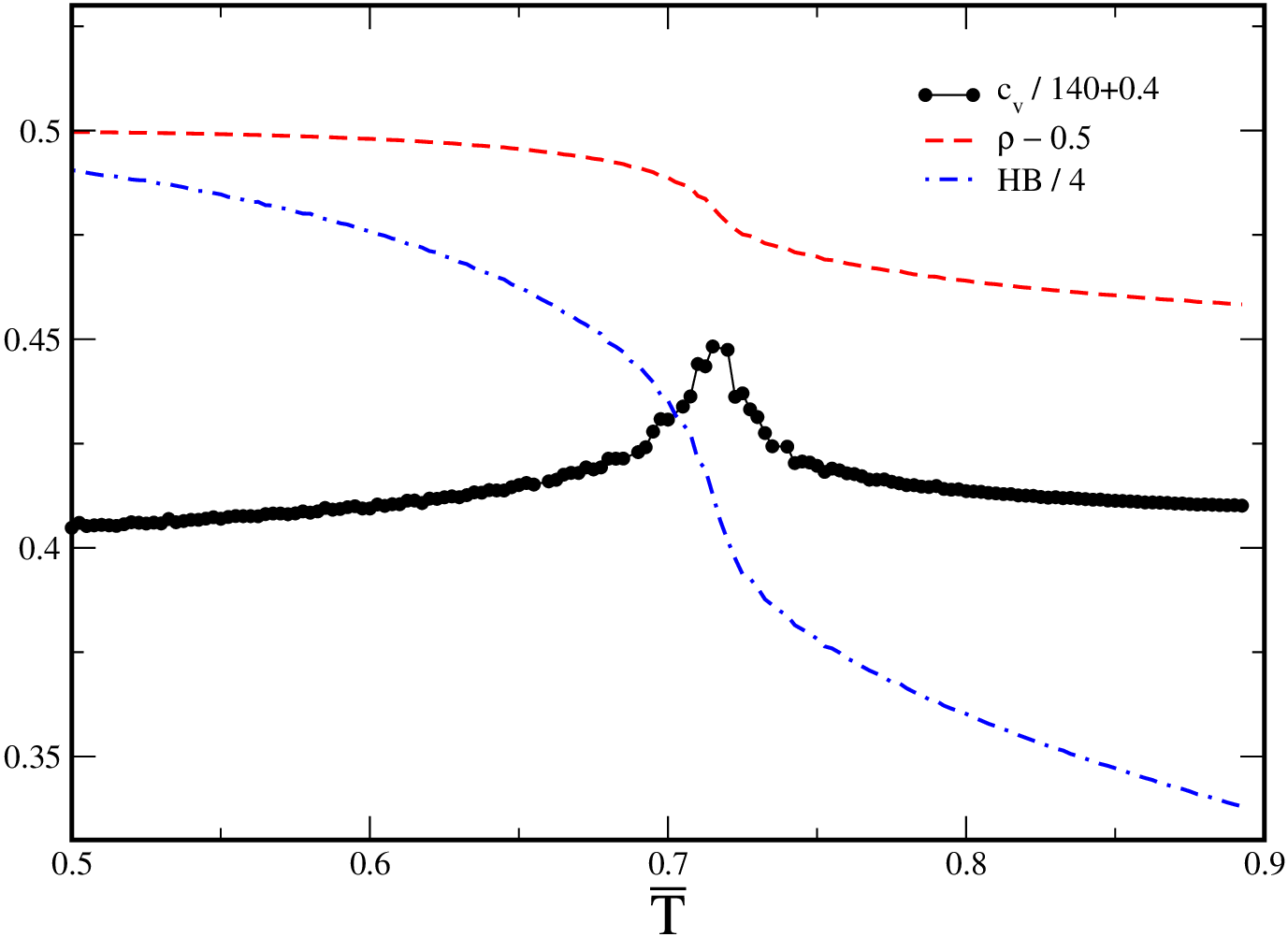}
\par\end{centering}
\caption{(Color on line) 
Density, specific heat and 
number of bonds per site for versus reduced temperature  $\overline{\mu}=2.5$} 
\label{dw_tau}
\end{figure}

\begin{figure}
\begin{centering}
\includegraphics[clip=true,scale=0.4]{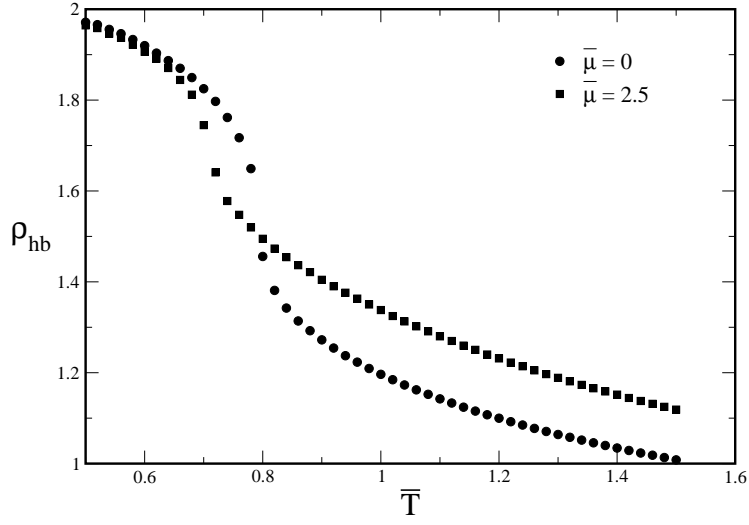} 
\par\end{centering}
\caption{ Density of hydrogen bonds as a  function of the reduced
temperature. 
For $\overline{\mu} = 2.5$ a smooth 
change in $\rho_{hb}$ is observed  while for
 $\overline{\mu} = 0$  $\rho_{hb}$ has
a more abrupt change characterizing a continuous transition.    }
\label{pontes} 
\end{figure}

\begin{figure}
\begin{centering}
\includegraphics[clip=true,scale=0.4]{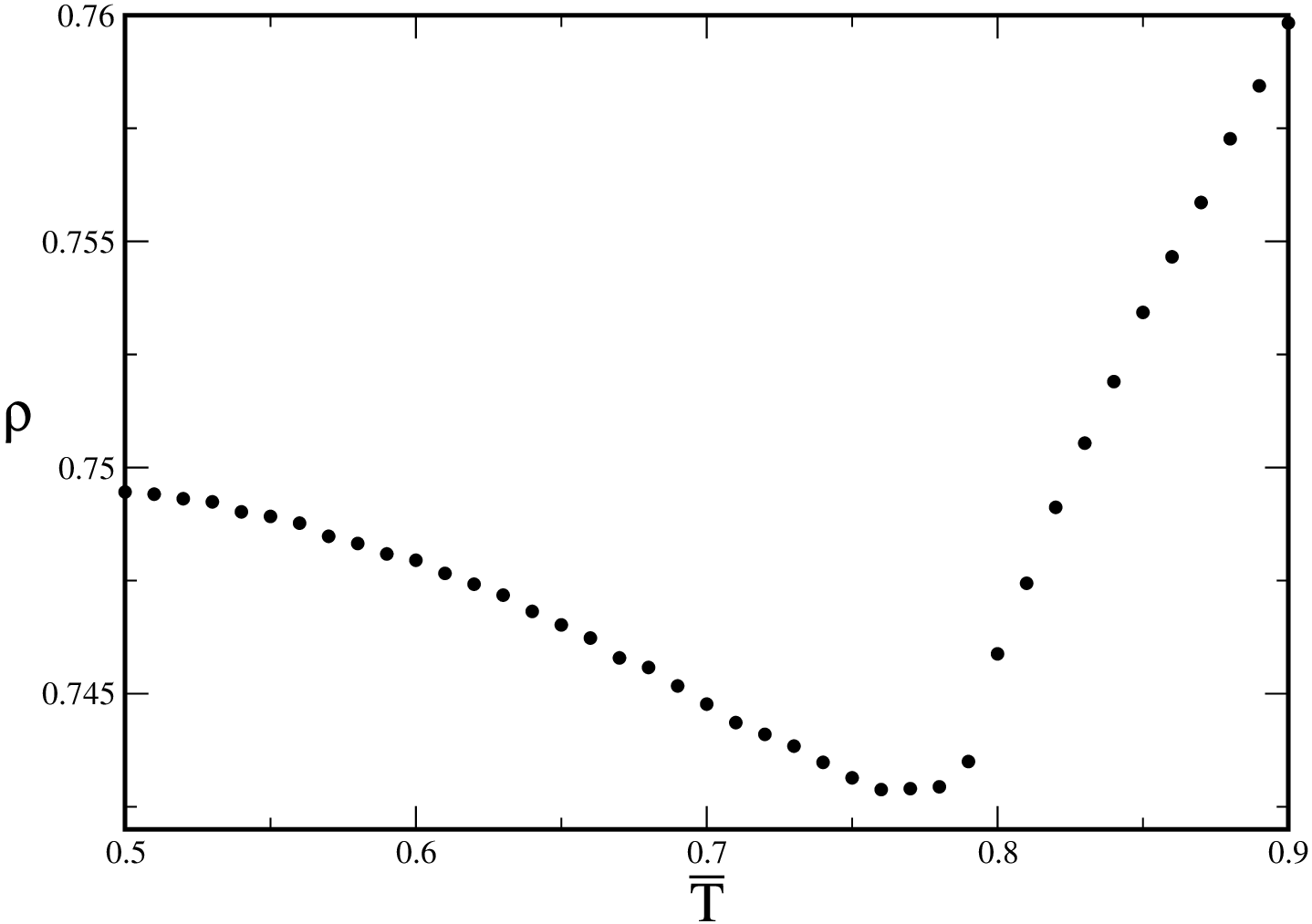} 
\par\end{centering}

\caption{ Density  as a  function of the reduced
temperature. 
for $\overline{\mu} = 0$.    }
\label{dw_mu_0} 
\end{figure}

\begin{figure}
\begin{centering}
\includegraphics[clip=true,scale=0.4]{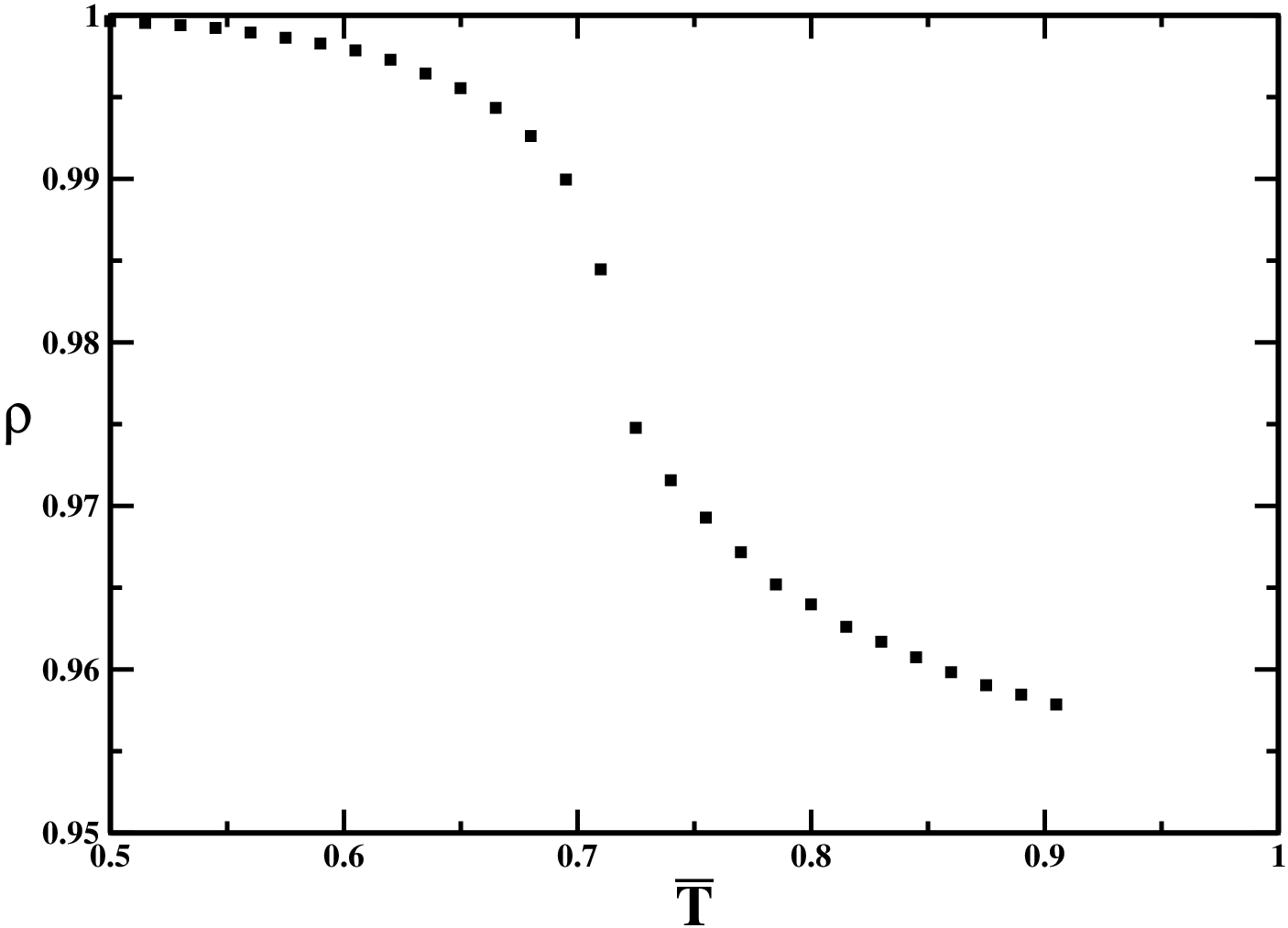} 
\par\end{centering}

\caption{ Density  as a  function of the reduced
temperature. 
for $\overline{\mu} = 2.5$.    }
\label{dw_mu_2.5} 
\end{figure}

In the attempt to understand the differences between the two transitions, it
is important to stablish 
what is the structural difference between the LDL, HDL, the high 
and the low densities 
fluid phases.
 To answer this question it is necessary to
establish a measure of
how structured is the liquid.
We adopt the number of hydrogen bonds per particle, $\rho_{hb}$, and
its correlation with particle density $\rho$ as
such a measure. 
 Fig.(\ref{dw_lambda}) shows that, as
temperature is decreased towards the specific heat peak position,
 hydrogen bond density increases,
while particle density decreases. This is indicative that bonding is
accompanied by particles abandoning the lattice.  
On the other hand, in the case of the $\tau$ line, both density 
and number of bonding particles increase as the temperature
decrease towards the peak temperature, as shown in Fig.(\ref{dw_tau}). 
Thus, in this case, bonding and lattice filling  occur simultaneously.
Fig.(\ref{pontes}) compares the behavior of bond density
$\rho_{hb}$ over the two transitions, while Figs. (\ref{dw_mu_0}) and 
(\ref{dw_mu_2.5}) illustrate the difference in density behavior.
In the case of the $\lambda$ transition, as temperature decreases towards the LDL phase, the density,
 shown in Fig.(\ref{dw_mu_0}), 
decreases drastically . At the transition,
the system orders itself by forming hydrogen bonds and
by releasing nonbonded particles. For temperatures below the transition,
the density increases. As for the $\tau$ line, 
density $\rho$, shown
in  Fig.(\ref{dw_mu_2.5}),increases smoothly as temperature is lowered.

%%%%%%%%%%%%%%%%%%%%%%%%%%%%%%%%%%%%%%%%%%%%%%%%%%%%%%%%%%%%%
%%%%%%%%%%%%%%%%%%%%%%%%%%%%%%%%%%%%%%%%%%%%%%%%%%%%%%%%%%%%%
%%%%%%%%%%%%%%%%%%%%%%%%%%%%%%%%%%%%%%%%%%%%%%%%%%%%%%%%%%%%%
%%%%%%%%%%%%%%%%%%%%%%%%%%%%%%%%%%%%%%%%%%%%%%%%%%%%%%%%%%%%%

%%%%%%%%%%%%%%%%%%%%%%%%%%%%%%%%%%%%%%%%%%%%%%%%%%%%%%%%%%%%%
%%%%%%%%%%%%%%%%%%%%%%%%%%%%%%%%%%%%%%%%%%%%%%%%%%%%%%%%%%%%%
%%%%%%%%%%%%%%%%%%%%%%%%%%%%%%%%%%%%%%%%%%%%%%%%%%%%%%%%%%%%%
%%%%%%%%%%%%%%%%%%%%%%%%%%%%%%%%%%%%%%%%%%%%%%%%%%%%%%%%%%%%%

%%%%%%%%%%%%%%%%%%%%%%%%%%%%%%%%%%%%%%%%%%%%%%%%%%%%%%%%%%%%%
%%%%%%%%%%%%%%%%%%%%%%%%%%%%%%%%%%%%%%%%%%%%%%%%%%%%%%%%%%%%%
%%%%%%%%%%%%%%%%%%%%%%%%%%%%%%%%%%%%%%%%%%%%%%%%%%%%%%%%%%%%%
%%%%%%%%%%%%%%%%%%%%%%%%%%%%%%%%%%%%%%%%%%%%%%%%%%%%%%%%%%%%%

%%%%%%%%%%%%%%%%%%%%%%%%%%%%%%%%%%%%%%%%%%%%%%%%%%%%%%%%%%%%%
%%%%%%%%%%%%%%%%%%%%%%%%%%%%%%%%%%%%%%%%%%%%%%%%%%%%%%%%%%%%%
%%%%%%%%%%%%%%%%%%%%%%%%%%%%%%%%%%%%%%%%%%%%%%%%%%%%%%%%%%%%%
%%%%%%%%%%%%%%%%%%%%%%%%%%%%%%%%%%%%%%%%%%%%%%%%%%%%%%%%%%%%%
%%%%%%%%%%%%%%%%%%%%%%%%%%%%%%%%%%%%%%%%%%%%%%%%%%%%%%%%%%%%%

%%%%%%%%%%%%%%%%%%%%%%%%%%%%%%%%%%%%%%%%%%%%%%%%%%%%%%%%%%%%%
%%%%%%%%%%%%%%%%%%%%%%%%%%%%%%%%%%%%%%%%%%%%%%%%%%%%%%%%%%%%%

%%%%%%%%%%%%%%%%%%%%%%%%%%%%%%%%%%%%%%%%%%%%%%%%%%%%%%%%%%%%%
%%%%%%%%%%%%%%%%%%%%%%%%%%%%%%%%%%%%%%%%%%%%%%%%%%%%%%%%%%%%%

%%%%%%%%%%%%%%%%%%%%%%%%%%%%%%%%%%%%%%%%%%%%%%%%%%%%%%%%%%%%%
%%%%%%%%%%%%%%%%%%%%%%%%%%%%%%%%%%%%%%%%%%%%%%%%%%%%%%%%%%%%%
\begin{figure}
\begin{centering}
\includegraphics[clip=true,scale=0.75]{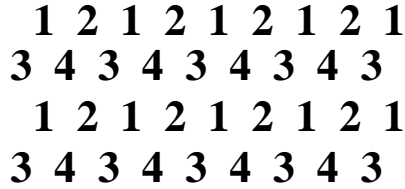} 
\par\end{centering}

\caption{ The lattice is divided into four sublattices  }
\label{sublattices} 
\end{figure}

%%%%%%%%%%%%%%%%%%%%%%%%%%%%%%%%%%%%%%%%%%%%%%%%%%%%%%%%%%%%%
%%%%%%%%%%%%%%%%%%%%%%%%%%%%%%%%%%%%%%%%%%%%%%%%%%%%%%%%%%%%%
\begin{figure}
\begin{centering}
\includegraphics[clip=true,scale=0.4]{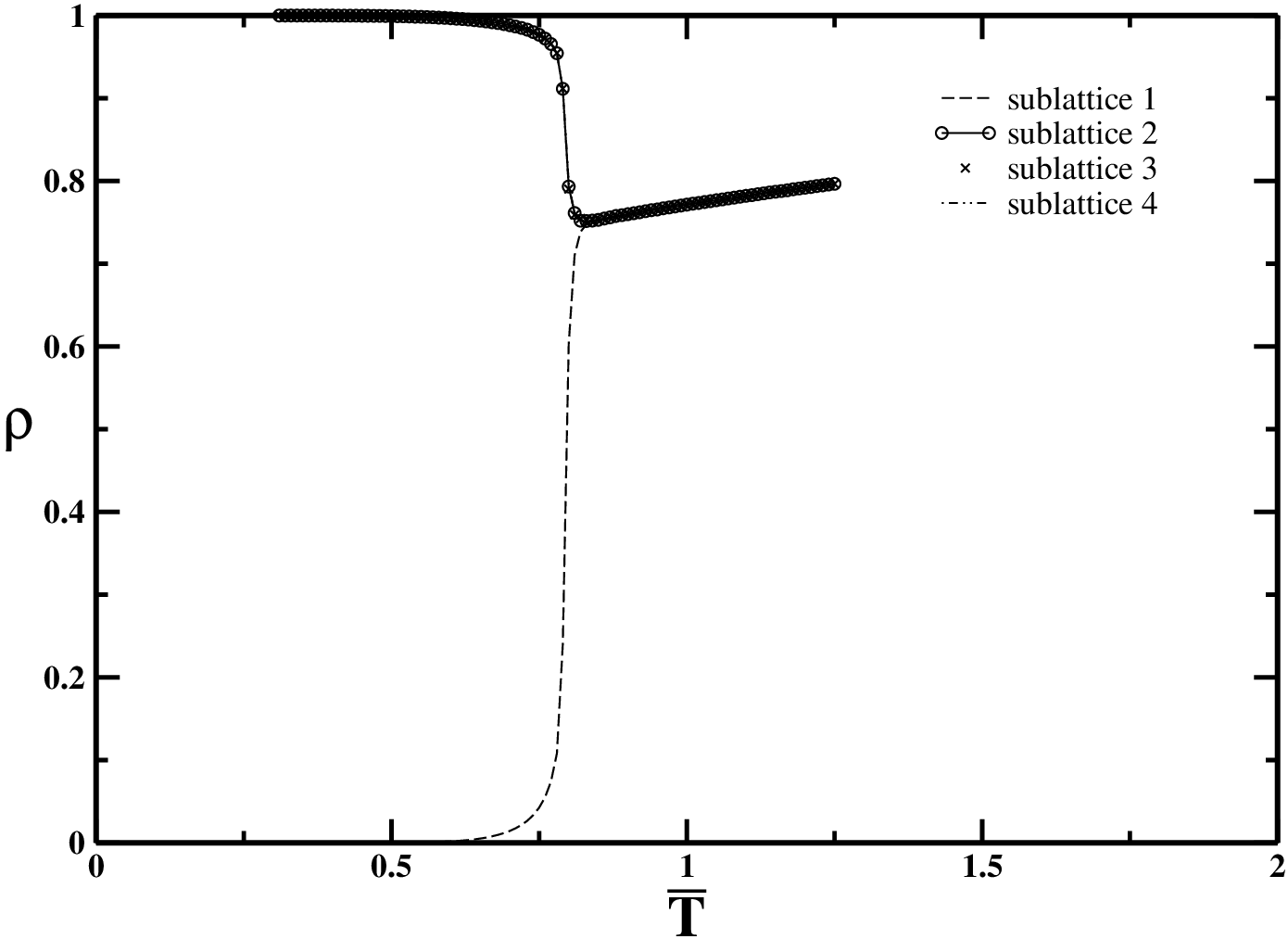} 
\par\end{centering}

\caption{Density of each one of the four sublattices 
as a function of temperature for $\overline{\mu}=0$. }
\label{rho_mu_0} 
\end{figure}

%%%%%%%%%%%%%%%%%%%%%%%%%%%%%%%%%%%%%%%%%%%%%%%%%%%%%%%%%%%%%

%%%%%%%%%%%%%%%%%%%%%%%%%%%%%%%%%%%%%%%%%%%%%%%%%%%%%%%%%%%%%
\begin{figure}
\begin{centering}
\includegraphics[clip=true,scale=0.4]{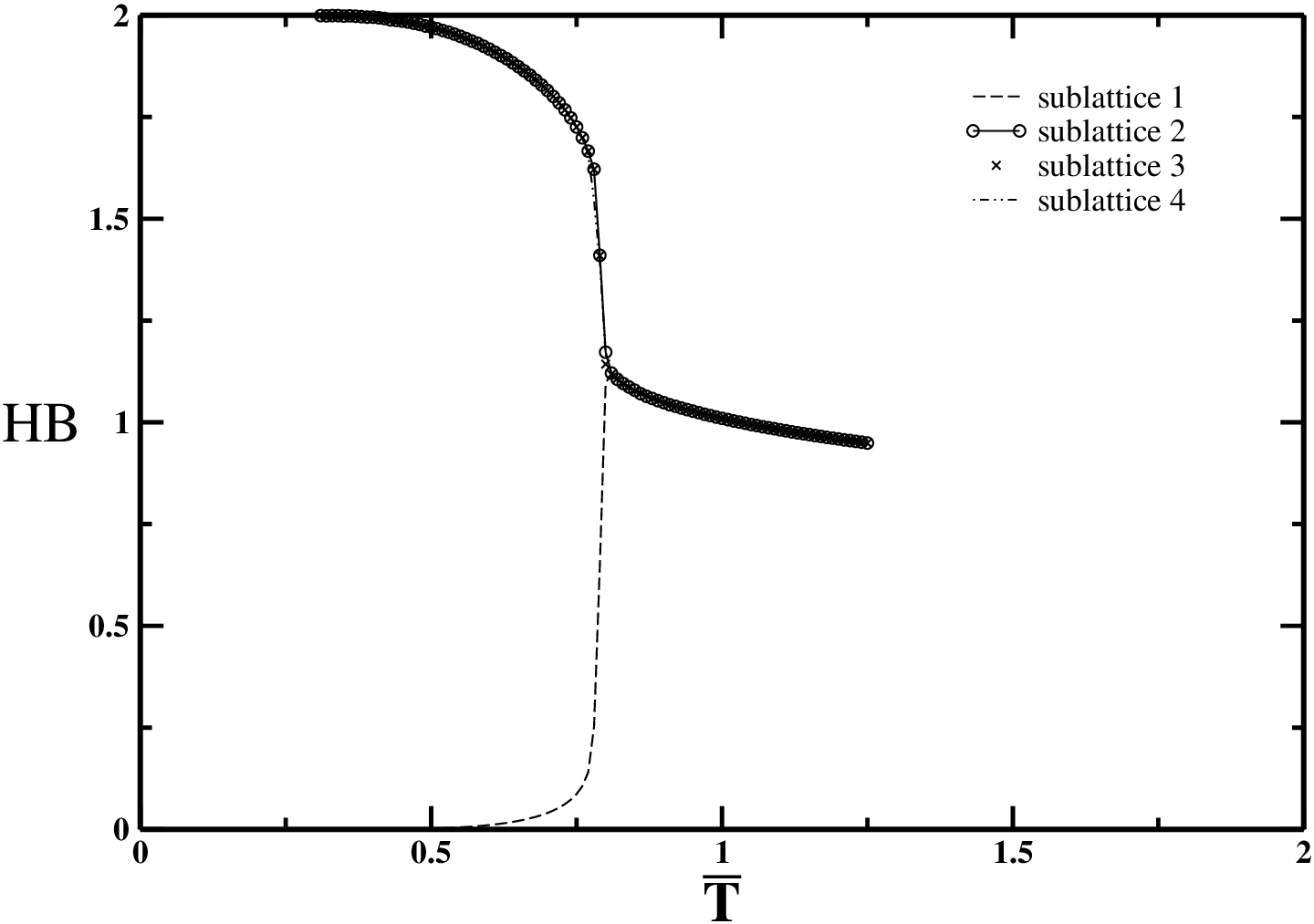} 
\par\end{centering}

\caption{Hydrogen bonds per site of each one of the four sublattices 
as a function of temperature for $\overline{\mu}=0$. }
\label{hb_mu_0} 
\end{figure}

%%%%%%%%%%%%%%%%%%%%%%%%%%%%%%%%%%%%%%%%%%%%%%%%%%%%%%%%%%%%%

%%%%%%%%%%%%%%%%%%%%%%%%%%%%%%%%%%%%%%%%%%%%%%%%%%%%%%%%%%%%%
\begin{figure}
\begin{centering}
\includegraphics[clip=true,scale=0.4]{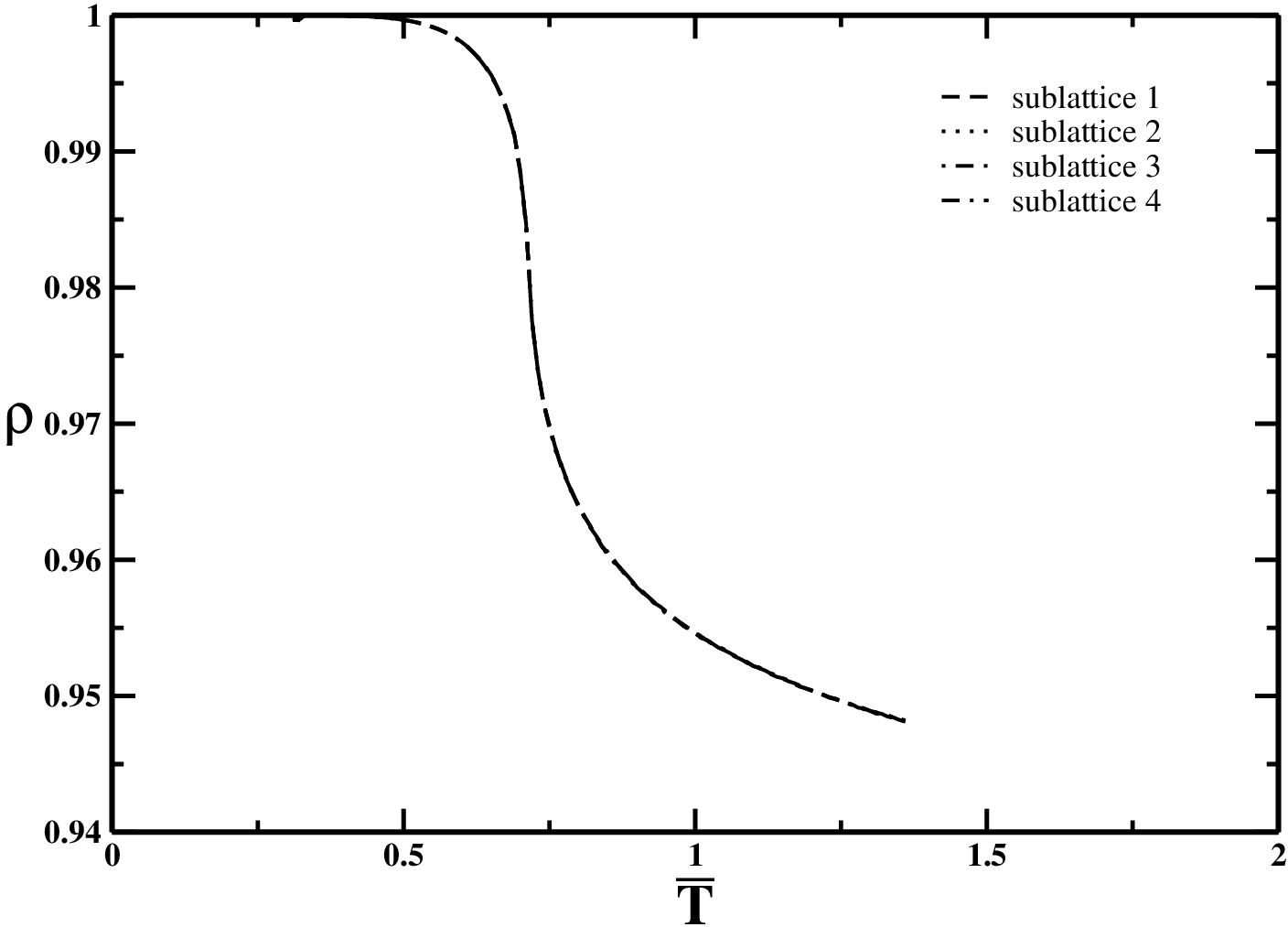} 
\par\end{centering}

\caption{Density of each one of the four sublattices 
as a function of temperature for $\overline{\mu}=2.5$. }
\label{rho_mu_2.5} 
\end{figure}

%%%%%%%%%%%%%%%%%%%%%%%%%%%%%%%%%%%%%%%%%%%%%%%%%%%%%%%%%%%%%

%%%%%%%%%%%%%%%%%%%%%%%%%%%%%%%%%%%%%%%%%%%%%%%%%%%%%%%%%%%%%
\begin{figure}
\begin{centering}
\includegraphics[clip=true,scale=0.4]{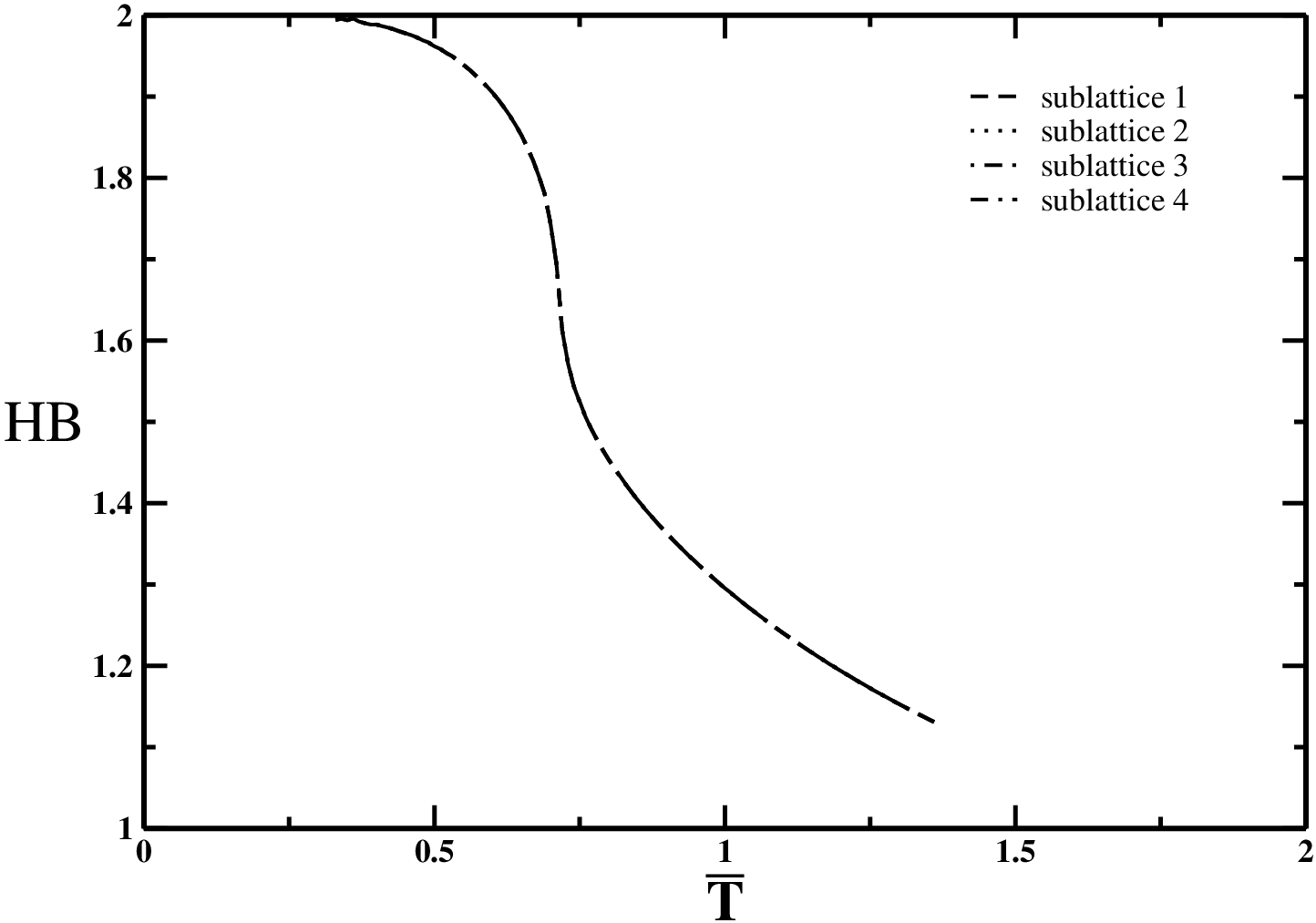} 
\par\end{centering}

\caption{Hydrogen bonds per site of each one of the four sublattices 
as a function of temperature for $\overline{\mu}=2.5$. }
\label{hb_mu_2.5} 
\end{figure}

%%%%%%%%%%%%%%%%%%%%%%%%%%%%%%%%%%%%%%%%%%%%%%%%%%%%%%%%%%%%%

A closer look  of this idea is possible if one examines the behavior of the
two densities on different sublattices. Inspection of Fig.(\ref{phases}) suggests dividing the lattice
into four sublattices, as illustrated in 
 Fig.(\ref{sublattices}). Figs.(\ref{rho_mu_0}) and (\ref{hb_mu_2.5}) display
sublattice density variations and how the number of  hydrogen bonds changes
with temperature, in the critical region, for $\overline{\mu}=0$. 
A clear critical transition is seen, in which one sublattice 
becomes empty, while the other three get ordered.   Figs.(\ref{rho_mu_2.5}) and (\ref{hb_mu_2.5})
illustrate density variations and the number of hydrogen bonds  of the four sublattices
with temperature, in the $\tau$ transition region, $\overline{\mu}=2.5$. In
this case, both densities of the four sublattices change smoothly.

Analysis of the sublattice data shows that
$\lambda$ is an order-disorder transition 
in which,  as the temperature
is decreased from the 
disordered fluid phase,  bonds are formed, 
while nonbonded sites  become empty.   This
critical $\lambda$ line joins the two coexistence 
lines at a tricritical point, $T_{c1}$, and a bicritical point, $T_{c2}$.

%%%%%%%%%%%%%%%%%%%%%%%%%%%%%%%%%%%%%%%%%
%%%%%%%%%%%%%%%%%%%%%%%%%%%%%%%%%%
\section{\label{sec4} Dynamics}
%%%%%%%%%%%%%%%%%%%%%%%%%%%%%%%%%%
%%%%%%%%%%%%%%%%%%%%%%%%%%%%%%%%%%%%%%%%%

In order to 
quantify mobility in
supercooled liquids, the concept
of  fragility  was introduced by Angell \cite{An97}.
Analyzing 
relaxation as a
function of temperature, liquids are classified as \emph{strong}, when 
relaxation follows 
an Arrhenius law, or \emph{fragile}, when the relaxation follows a 
non-Arrhenius law. 
Strong liquids present structure that is preserved when temperature is 
increased, 
whereas in fragile liquids this structure is easily broken, as 
temperature increases. 

Within the framework of the  Adam-Gibbs theory \cite{Ad65}, viscous 
liquids are described as being made of  clusters that 
rearrange cooperatively in order to pass through the free energy barrier.
Consequently, diffusion depends on this
cooperative rearrangement of the clusters through equation
%%%%%%%%%%%%%%%%%%%%%%%%%%%%%%%%%%%%%%%%%%%%%%%%%%%%%%%%%%%%%
\begin{equation}
D = D_{0}\;exp\left(\frac{C\;\Delta\mu}{TS_{c}}\right)\;\;,\label{adam}
\end{equation}
%%%%%%%%%%%%%%%%%%%%%%%%%%%%%%%%%%%%%%%%%%%%%%%%%%%%%%%%%%%%%
 for the diffusion constant $D$. Here $D_{0}$ and $C$ are constants, $\Delta\mu$ is the free energy barrier 
which the clusters have
to overcome. $S_{c}$ is the configurational entropy, given by
%%%%%%%%%%%%%%%%%%%%%%%%%%%%%%%%%%%%%%%%%%%%%%%%%%%%%%%%%%%%%
\begin{equation}
S_{c}(T) = \int_{T_{K}}^{T} \left(\frac{\Delta C_{p}}{T}\right) dT\;\;, 
\label{entro}
\end{equation}
%%%%%%%%%%%%%%%%%%%%%%%%%%%%%%%%%%%%%%%%%%%%%%%%%%%%%%%%%%%%%
that describes how the structure of the 
liquid changes with temperature. 
In Eq.(\ref{entro})  $T_{K}$ is the Kauzmann temperature \cite{An97}(for 
which $S_{c}(T_{K}) = 0$)
and $\Delta c_{p}$ is the difference in specific heat between the crystal 
 and the liquid configurations, at temperature $T$. 
If $S_{c}$,  
is temperature independent, the diffusion follows an Arrhenius law,
the liquid is very structured
 and the system is a strong liquid. 
If the configurational entropy depends on temperature, 
$S_{c} = \Delta C_{p} ln\;T/T_{k}$, Eq.(\ref{adam}) becomes
a Vogel-Fulcher equation, the liquid 
is not structured and is classified as a fragile liquid.

Now we investigate the dynamic properties on
crossing  the $\lambda$ and the $\tau$ lines, at constant chemical
potential (see Fig. (\ref{mu-t})), by analysing behavior of model
diffusivity.
In order to compute diffusion coefficient we first equilibrate the system 
at fixed chemical potential and temperature.  In equilibrium
this system has $n$ particles.  
Starting from this equilibrium configuration at a time $t=0$, each
one of these  $n$
particles is allowed to move to an empty neighbor site
randomly chosen. The movement
is accepted if the total energy of the system 
is reduced, otherwise it is accepted with
a probability $\exp(\Delta E /k_BT)$ where $\Delta E$ is 
the difference between the energy of the system after and before the
movement. After repeating this procedure  $nt$ times, the mean 
square displacement per particle  at a time $t$
is given by
%%%%%%%%%%%%%%%%%%%%%%%%%%%%%%%%%%%%%%%%%%
\begin{equation}
\langle \Delta r(t)^2 \rangle =\langle\left(\textbf{r}(t)-
\textbf{r}(0)\right)^2 \rangle\;\;,
\label{r2}
\end{equation}
%%%%%%%%%%%%%%%%%%%%%%%%%%%%%%%%%%%%%%%%%%%
where  $\textbf{r}(0)$ is the particle
position at the initial time and  $\textbf{r}(t)$ is the 
particle position at a time $t$. In Eq.~(\ref{D}), the average 
is taken over all particles and over 
different initial configurations.
The diffusion coefficient is then obtained from Einstein's relation
%%%%%%%%%%%%%%%%%%%%%%%%%%%%%%%%%%%%%%%%%%%%%%%
\begin{equation}
D=\lim_{t\rightarrow \infty}\frac{\langle \Delta r(t)^2 \rangle}{4t}\; .
\end{equation}
%%%%%%%%%%%%%%%%%%%%%%%%%%%%%%%%%%%%%%%%%%%%%%%
Since the time is measured in Monte Carlo time steps  and 
the distance in number of lattice distance, a dimensionless 
diffusion coefficient is defined as
%%%%%%%%%%%%%%%%%%%%%%%%%%%%%%%%%%%%%%%%%%%%%%%
\begin{equation}
\overline{D}=\lim_{t\rightarrow \infty}\frac{\langle \Delta
\overline{r}(t)^2 \rangle}{4\overline{t}}\; .
\label{D}
\end{equation}
%%%%%%%%%%%%%%%%%%%%%%%%%%%%%%%%%%%%%%%%%%%%%%%
where $\overline{r}=r/a$ and $a$ is the distance 
between two neighbor sites and 
$\overline{t}=t/t_{MC}$
is the time in Monte Carlo steps.

\begin{figure}
 \centering
 \subfigure[Diffusion as a function of 
      temperature for $\overline{\mu} = -0.25$.] 
% caption for subfigure a
 {
     \label{smu-0.25}
     \includegraphics[clip=true,scale=0.30]{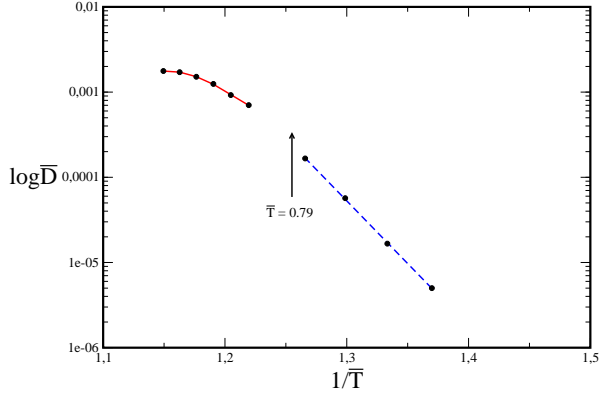}
     }
 \subfigure [Diffusion in function of 
     temperature for $\overline{\mu} = 0$.] 
% caption for subfigure b
 {
     \label{smu-0}
     \includegraphics[clip=true,scale=0.30]{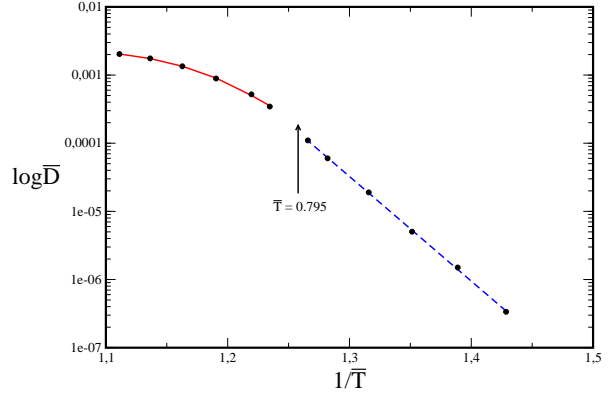}
     
 }
 \subfigure [ Diffusion in function of 
     temperature for $\overline{\mu} = 0.5$.] 
% caption for subfigure c
 {
     \label{smu-0.5}
     \includegraphics[clip=true,scale=0.30]{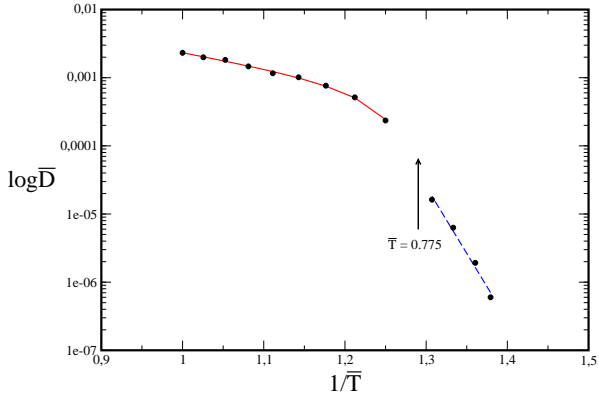}
 }
 \subfigure [ Diffusion in function of temperature for 
$\overline{\mu} = 1$.] 
% caption for subfigure b
 {
     \label{smu-1}
     \includegraphics[clip=true,scale=0.30]{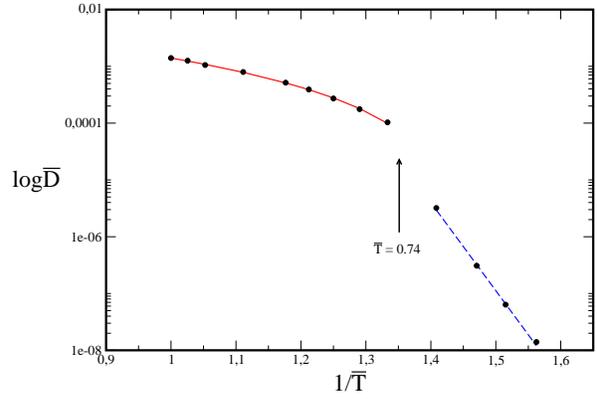}
 }

 \caption{(Color on line). The circles are diffusion coefficient 
measured in simulation,  
solid line is a cubic fit and dashed line is an exponential fit. In high temperatures system 
behaves as a fragile liquid 
following a non-Arrhenius law, while for low temperatures the 
system behaves like a
 strong liquid following an Arrhenius law.}
 \label{transition1} % caption for the whole figure
 \end{figure}

Figs.(\ref{smu-0.25}) - (\ref{smu-1}) illustrate the 
behavior of the diffusion constant $D$ with the inverse
of the reduced temperature $1/\overline{T}$, 
for fixed values of the reduced chemical potentials ($\overline{\mu}=-0.25,\;0,\;0.5,\;1 $)
At higher temperatures, diffusivity follows a non-Arrhenius trend,
namely
%%%%%%%%%%%%%%%%%%%%%%%%%%%%%%%%%%%%%%%%%%%%%%%%%%%%%%%%%%%%%
\begin{equation}
y = A_{0} + A_{1} x + A_{2} x^{2} + A_{3} x^{3}  \;\;
\label{cubic}
\end{equation}
%%%%%%%%%%%%%%%%%%%%%%%%%%%%%%%%%%%%%%%%%%%%%%%%%%%%%%%%%%%%%
indicating that the low density disordered fluid phase is a fragile liquid. At lower 
temperatures, diffusivity displays Arrhenius behavior,
given by
%%%%%%%%%%%%%%%%%%%%%%%%%%%%%%%%%%%%%%%%%%%%%%%%%%%%%%%%%%%%%
\begin{equation}
y = A_{0}\; exp\left(-\frac{A_{1}}{x}\right) \;\;
\label{arrhenius}
\end{equation}
%%%%%%%%%%%%%%%%%%%%%%%%%%%%%%%%%%%%%%%%%%%%%%%%%%%%%%%%%%%%%
thus characterizing the
low density ordered liquid phase as a strong liquid. 
$A_{i}$ are fitting parameters in both equations.
%%%%%%%%%%%%%%%%%%%%%%%%%%%%%%%%%%%%%%%%%%%%%%%%%%%%%%%%%%%%%
%%%%%%%%%%%%%%%%%%%%%%%%%%%%%%%%%%%%%%%%%%%%%%%%%%%%%%%%%%%%%
%%%%%%%%%%%%%%%%%%%%%%%%%%%%%%%%%%%%%%%%%%%%%%%%%%%%%%%%%%%%%
 \begin{figure}
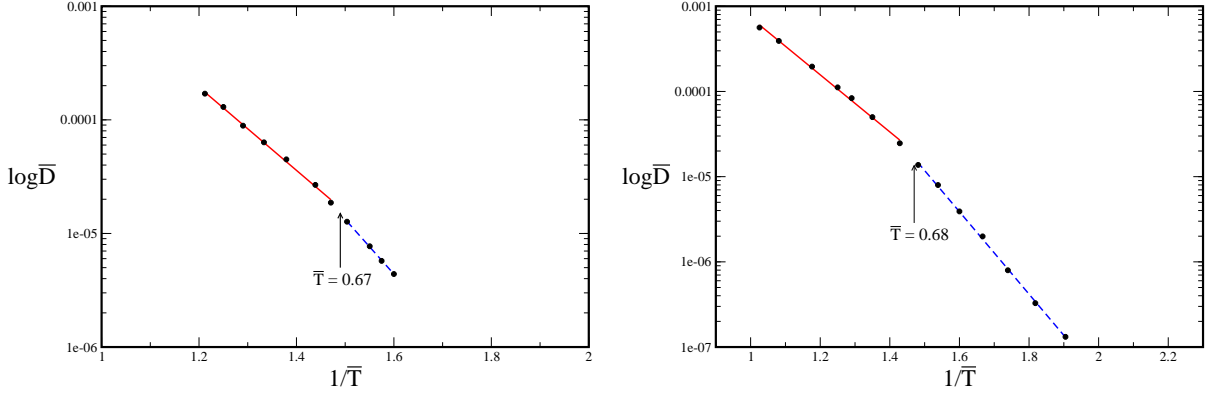
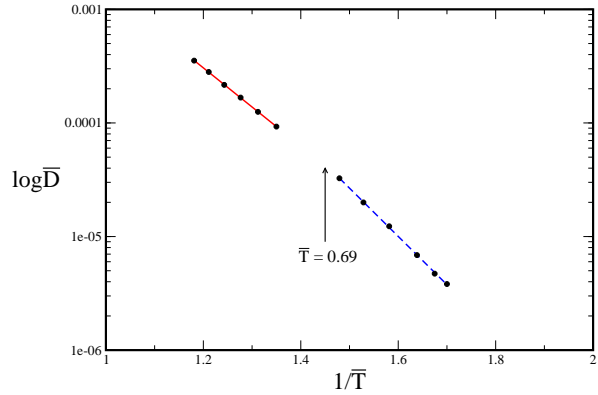

 \centering
 \subfigure [Diffusion in function of 
temperature for $\overline{\mu} = 1.75$.] % caption for subfigure a
 {
     \label{smu-1.75}
     \includegraphics[clip=true,scale=0.30]{smu_1.75.eps}
     
 }
 \subfigure [Diffusion in function of 
temperature for $\overline{\mu} = 1.85$.] % caption for subfigure b
 {
     \label{smu-1.85}
     \includegraphics[clip=true,scale=0.30]{smu_1.85.eps}
 }
 \subfigure [Diffusion in function of 
temperature for $\overline{\mu} = 1.95$.] % caption for subfigure c
 {
     \label{smu-1.95}
     \includegraphics[clip=true,scale=0.30]{smu_1.95.eps}
 }

 \caption{(Color on line). Circles are diffusion coefficient measured 
in simulation and 
solid and dashed lines are two different exponential fits. In the region of $\tau$ line system 
behaves as a strong liquid in both 
sides. The difference between the two Arrhenius behaviors is the 
activation energy, that 
defines the slope of the curve.}
 \label{transition2} % caption for the whole figure
 \end{figure}
%%%%%%%%%%%%%%%%%%%%%%%%%%%%%%%%%%%%%%%%%%%%%%%%%%%%%%%%%%%%%

This change in dynamics over the critical $\lambda$ line occurs because the liquid is structurally 
different 
on both sides of the critical line.  In the low density disordered fluid phase, interstitial 
particles 
weaken the  hydrogen bonds and disrupt the network, so particles can rearrange 
fast and the process of diffusion is not energy activated. In the LDL phase, 
the 
network is fully 
developed, resulting in an ordered liquid, in which particles 
are "trapped", increasing 
relaxation time and characterizing this phase as a strong liquid, 
in which an energy activated 
diffusion process takes place.  This is the dynamic transition
observed when crossing a Widom line in ramp-like
models \cite{Xu05}\cite{Xu06}, which suggests that the dynamic
transition is not linked with the type of line but
with the structuring of the system if this happens with
or without a thermodynamic phase-transition.
The system becomes more
organized, as can be seen from the drastic change
in the density of the sublattices shown in Fig. (\ref{rho_mu_0})
with a smooth change in the total density. 

Since the HDL is also a structured phase, in principle 
a fragile-strong transition in the dynamics of diffusion
could also be expected 
on crossing the $\tau$ line.   However, this is 
not the case. Figs.(\ref{smu-1.75}) - (\ref{smu-1.95})
illustrate the behavior of the  diffusion constant as function of inverse
temperature,  $1/\overline{T}$, for
fixed chemical potentials $\overline{\mu} = 1.75, 1.85$ and $1.95$.
At higher temperatures and high chemical
potentials (or equivalently high densities), the 
fluid phase has an Arrhenius behavior and so it is
a strong liquid. At lower temperatures, the HDL phase 
also displays an Arrhenius behavior, and therefore is also
a strong liquid. The HDL phase and high density fluid
phases are both strong liquids
that differ in the activation energy.
 In resume, when the system crosses the $\tau$ line, we have
a dynamic transition, and a strong-strong crossover is observed.
In this case, the activation energy  of the HDL phase is higher than
the activation energy   of the 
high density fluid phase, indicating that the HDL phase is more ordered than the
high density fluid phase. Diffusion is lowered in the HDL phase because 
particles 
spend more 
time trying to rearrange, in comparison with the high density fluid phase. 

How can we explain the existence of 
a fragile-to-strong crossover  on the critical $\lambda$-line and 
a strong-to-strong transition on the $\tau$ line?
 The answer is given by the structure of the liquid, described in the
previous section. On crossing the $\lambda$-line, the hydrogen-bonded net breaks down abruptly (see Fig. \ref{hb_mu_0}),
while the $\tau$-line is accompanied by a much smoother melting of the h-bond network (see Fig.\ref{hb_mu_2.5} ).
The difference in the change of structure on crossing one and the other lines is even clearer if
one looks at sublattice densities. See Fig.(\ref{rho_mu_0}) and Fig.(\ref{rho_mu_2.5}).

%%%%%%%%%%%%%%%%%%%%%%%%%%%%%%%%%%%
%%%%%%%%%%%%%%%%%%%%%%%%%%%%%%%%%%%%
\section{\label{sec5} Conclusions}
%%%%%%%%%%%%%%%%%%%%%%%%%%%%%%%%%%%%
%%%%%%%%%%%%%%%%%%%%%%%%%%%%%%%%%%%%

In this paper we have analyzed equilibrium and dynamic 
properties of the Associating Lattice Gas Model, 
a lattice gas with  hydrogen bonds represented
 through ice variables. Competition between the filling
up of the lattice and the formation of an open four-bonded orientational 
structure leads to the presence of two liquid phases and a gas phase.
The coexistence lines  between the LDL and the gas phases, and between 
the LDL and HDL phases are connected by a critical $\lambda$-line. Besides the 
$\lambda$-line, a second one, the $\tau$-line, also emerges from 
the LDL-HDL coexistence line. This line is also identified
 by a peak in the specific heat.

The system undergoes two kinds of dynamic transitions: 
one fragile-to-strong (crossing the $\lambda$-line) and 
a strong-to-strong (crossing the $\tau$-line). Both dynamic 
transitions are related with structure of the system. 
In the fragile-to-strong case, the system change its 
structure drastically when cross $\lambda$-line, from 
a disordered structure in high temperatures to an ordered one 
at low temperatures. In the $\tau$-line region, 
the change in structure is more subtle, but again the system 
undergoes a structural change between two ordered phases.

Our results point out in the direction that criticality does
not necessarily means fragil-strong transtion. This change
is in fact related to the change of structure that in
the present case appears in two very different forms.

\subsection*{Acknowledgments}

This work was supported by the Brazilian science agencies CNPq,
FINEP, Capes  and Fapergs.

\end{document}